\documentclass[12pt]{iopart}

\usepackage{graphicx}
\usepackage{iopams}
\usepackage[compress]{cite}

\begin{document}

\title[Thermoelectric Power Measurement in Zero and Applied Magnetic Field]
{Experimental Setup for the Measurement of the Thermoelectric Power in Zero and Applied Magnetic Field}

\author{Eundeok Mun$^{1}$, Sergey L. Bud'ko$^{1}$, Milton S. Torikachvili$^{2}$, Paul C. Canfield$^{1}$}
\address{$^{1}$Ames Laboratory US DOE and Department of Physics and Astronomy, Iowa State University, Ames, IA 50011, USA}
\address{$^{2}$Department of Physics, San Diego State University, San Diego, California 92182-1233 USA}

\ead{canfield@ameslab.gov}
\begin{abstract}
An experimental setup was developed for the measurement of the thermoelectric power (TEP, Seebeck coefficient) in the temperature range from 2 to
350 K and magnetic fields up to 140 kOe. The system was built to fit in a commercial cryostat and is versatile, accurate and automated; using two
heaters and two thermometers increases the accuracy of the TEP measurement. High density data of temperature sweeps from 2 to 350 K can be
acquired in under 16 hours and high density data of isothermal field sweeps from 0 to 140 kOe can be obtained in under 2 hours. Calibrations for
the system have been performed on a platinum wire and Bi$_{2}$Sr$_{2}$CaCu$_{2}$O$_{8+\delta}$ high $T_{c}$ superconductors. The measured TEP of
phosphor-bronze (voltage lead wire) turns to be very small, where the absolute TEP value of phosphor-bronze wire is much less than 0.5 $\mu$V/K
below 80 K. For copper and platinum wires measured against to the phosphor-bronze wire, the agreement between measured results and the literature
data is good. To demonstrate the applied magnetic field response of the system, we report measurements of the TEP on single crystal samples of
LaAgSb$_{2}$ and CeAgSb$_{2}$ in fields up to 140 kOe.
\end{abstract}

\vspace{2pc} \noindent{\it Keywords}: thermoelectric power, measurement setup, calibration
\maketitle

\section{Introduction}
Since its discovery in 1821 by Thomas Johann Seebeck, relatively few studies of the magnetic field dependence of the thermoelectric power (TEP)
were carried out, mostly in pure metals \cite{Blatt}. However, over the past few decades, the magnetic field-dependent TEP studies of many
materials ranging from magnetic multilayers \cite{Sakurai} to high $T_{c}$ superconductors \cite{Wang}, to the electron-topological transition
\cite{Budko2} and to strongly correlated electron systems \cite{Sakurai1, Benz, Izawa} have been provided useful information. Intensive efforts
also have been made in the search for highly efficient thermoelectric materials. This being said, the measurement of the intrinsic TEP is
particularly difficult even in simple metals such as copper or gold. This is due to the small magnitude of TEP at low temperatures and its
sensitivity to the presence of small concentrations of impurities, where magnetic impurities can enhance the TEP below certain temperatures by
means of the Kondo effect \cite{Blatt}.

Few experimental details have been given in the literature concerning the measurement setups and the procedure for calibration of lead (as in
contacting the sample, not Pb) wires \cite{Burkov, Choi, Resel}. Detailed descriptions of the measurement techniques at low temperatures and high
magnetic fields can be found in Refs. \cite{Choi, Resel}. In this article, we describe the development of an experimental setup for TEP
measurement in a Quantum Design (QD), Physical Property Measurement System (PPMS). The PPMS sample puck provides both thermal and electrical
contacts to the sample. The merits of this technique are (i) it is easy to implement using two commercial, Cernox thin-film, resistance cryogenic
temperature sensors and two strain gauge heaters and (ii) it is easy to control the temperature and magnetic field of the system using the PPMS
platform. Using the PPMS temperature-magnetic field ($T-H$) environment and the two heaters and two thermometers, an alternating heating method
allows for measurements of the TEP of materials over a temperature range from 2 to 350 K and magnetic fields up to 140 kOe. The alternating
heating method we use improves the resolution by a factor of two and provides a reliable temperature gradient. For the measurement, the sample is
mounted directly between the two Cernox thermometers each of which is heated by a strain gauge heater with constant DC current. An important
component of this technique involves the use of phosphor-bronze lead wires to reduce the background TEP and magneto-thermoelectric power (MTEP)
associated with the lead wires.

\section{Experimental Setup}
In this section we will describe our specific sample holder (sample stage) and explain data acquisition process. This measurement setup was
designed to fit PPMS cryostat used to control the temperature and magnetic field of the system. All instruments (current sources, voltmeters,
switch system and PPMS) were controlled by National Instruments LabVIEW software. The sample holder can be easily modified and adapted to other
cryogenic systems, including those with higher magnetic fields and lower temperatures.

\subsection{Sample Holder}
\begin{figure*}
\centering
\includegraphics[width=1\linewidth]{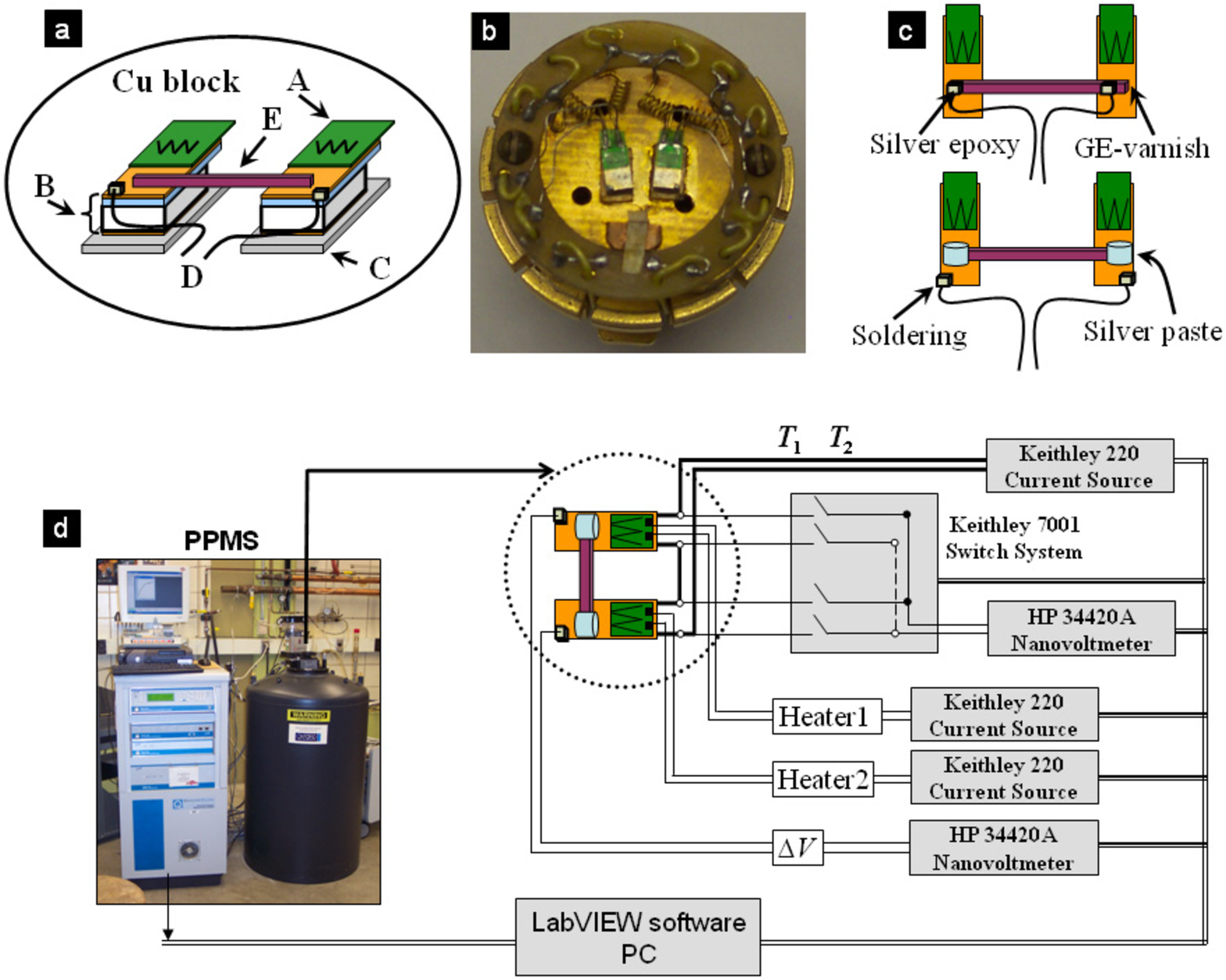}
\caption{(a) Schematic diagram of sample stages.  A: Strain gauges for heater, B: Thermometers (Cernox), C: G-10 for thermal insulation from heat
sink, D: Voltage probe wires, E: Sample. (b) A photo of the measurement cell. (c) Sample mounting method using GE-varnish (top) and silver paste
(bottom). (d) Block diagram of measurement system. The system temperature and magnetic field is controlled by PPMS. All instruments shown in the
block diagram including PPMS is operated by LabVIEW software. The details of the use of the instruments are explained in the text.}\label{holder}
\end{figure*}

Figures \ref{holder} (a) and (b) show a schematic diagram of the sample stage built on the PPMS sample puck and a photograph of actual sample
stage. The magnetic field is applied perpendicular to the plane of the heaters, thermometers and puck platform. Two sample stages are attached to
a circular copper heat sink positioned on the 23 mm diameter PPMS sample puck that, when in use, is shielded by a gold plated copper cap (not
shown). We use Cernox sensors (CX-1050-SD package) as thermometers that provide high sensitivity at low temperatures, good sensitivity over a
broad range and low magnetic field-induced errors. The dimensions of this package (1.9$\times$1.1$\times$3.2 mm$^{3}$) are large enough to attach
a heater and sample simultaneously to the package surface. Strain gauges (heaters), 0.2 $\times$ 1.4 mm$^{2}$ and typically $R\sim$ 120 $\Omega$,
are glued to the top of the Cernox thermometers using Stycast 1266 epoxy. In order to insure thermal isolation, the heat sink (PPMS puck) and the
sample stage was separated by a thin (1 mm thickness) G-10 plate. This G-10 plate was glued on the bottom of the Cernox thermometer using the
Stycast 1266 epoxy. From several test runs we observed that the two Cernox wires and two heater wires provided enough cooling power to the sample
stage since the strain gauge and Cernox each have low thermal mass. Each sample stage including heater, thermometer and G-10 plate was glued to
the copper heat sink with GE 7301 varnish, so that it was easy to remove by dissolving the GE-varnish with ethanol. Because of the constraint of
the PPMS sample puck, the distance between two stages can be varied from $\sim$1.5 mm to $\sim$ 6 mm. Large flexibility with respect to the
sample size can therefore be gained since the precise configuration of the thermal stage can be easily adjusted. If the sample length is smaller
than 1.5 mm, it is hard to establish a temperature difference ($\Delta T$) because both thermal stages are isolated from the heat sink.
Typically, samples with length varying from 2 to 7 mm can be measured. All wires on the measurement cell are thermally anchored to the heat sink.
The TEP measurement was made with the PPMS operating in the high vacuum mode with pressure $\sim10^{-5}$ torr.

For mounting the sample, and measuring the voltage, two different configurations were tested (Fig. \ref{holder} (c)). First, samples were mounted
on the two sample stages with GE-varnish. The voltage difference $\Delta V$ is measured using 25 $\mu$m diameter copper wire or phosphor-bronze
wire attached to the sample using silver epoxy as shown in the top of Fig. \ref{holder} (c). Alternatively, samples were directly mounted to the
sample stages using DuPont 4929N silver paste. The silver paste provides good thermal and electrical contact between the sample and the gold
plated layer on the surface of the Cernox package (bottom of Fig. \ref{holder} (c)). The copper wire or phosphor-bronze lead wire is soldered to
this gold plated layer. In this case the voltage difference is obtained by measuring the voltage difference between two sample stages. Since the
data was taken in a steady state, by assuming the temperature of the gold layer is the same as silver paste, the TEP contribution of the sample
stage can be ignored. Since the silver paste can be dissolved in hexyl acetate, the sample can be easily detached by carefully adding small amount
of this solvent without degrading Stycast or GE-varnish. We ran several test measurements to compare thermal coupling between sample and
thermometers by using silver paste and GE-varnish. We found it to be essentially the same for both cases. In general the TEP measurement was
performed with the later, silver paste configuration, because the sample mounting and removal were easier than GE-varnish. The GE-varnish
configuration is preferred mainly when good electrical contact between the sample and the gold layer of the thermometer with silver paste can not
be established. For example, when we measure the TEP of the Bi$_{2}$Sr$_{2}$CaCu$_{2}$O$_{8+\delta}$ (Bi2212) high $T_{c}$ samples for calibration
it was hard to get good electrical contact (see next section).

\subsection{Determination of $\Delta T$, $\Delta V$, $T_{av}$ and $S$}
A block diagram of the TEP measurement is shown in Fig. \ref{holder} (d). Since the PPMS sample puck provides only 12 wires, they had to be used
frugally: Six wires total were used for the two Cernox sensors, which were connected in series, four wires were used for the heaters (2 each),
and two wires were used for the TEP voltage. The resistance of each Cernox is measured with a Hewlett Packard 34420A nanovoltmeter via a Keithley
7001 switch system with a Keithley 7059 low voltage scanner card. The current was supplied to the Cernox thermometers by a Keithley 220
programmable current source. A temperature difference ($\Delta T$) across the sample was established by applying a DC current with two Keithley
220 programmable current source alternately through one of the strain gauges at a time, while the voltage difference ($\Delta V$) across the
sample was monitored independently with a Hewlett Packard 34420A nanovoltmeter.

\begin{figure*}
\centering
\includegraphics[width=0.5\linewidth]{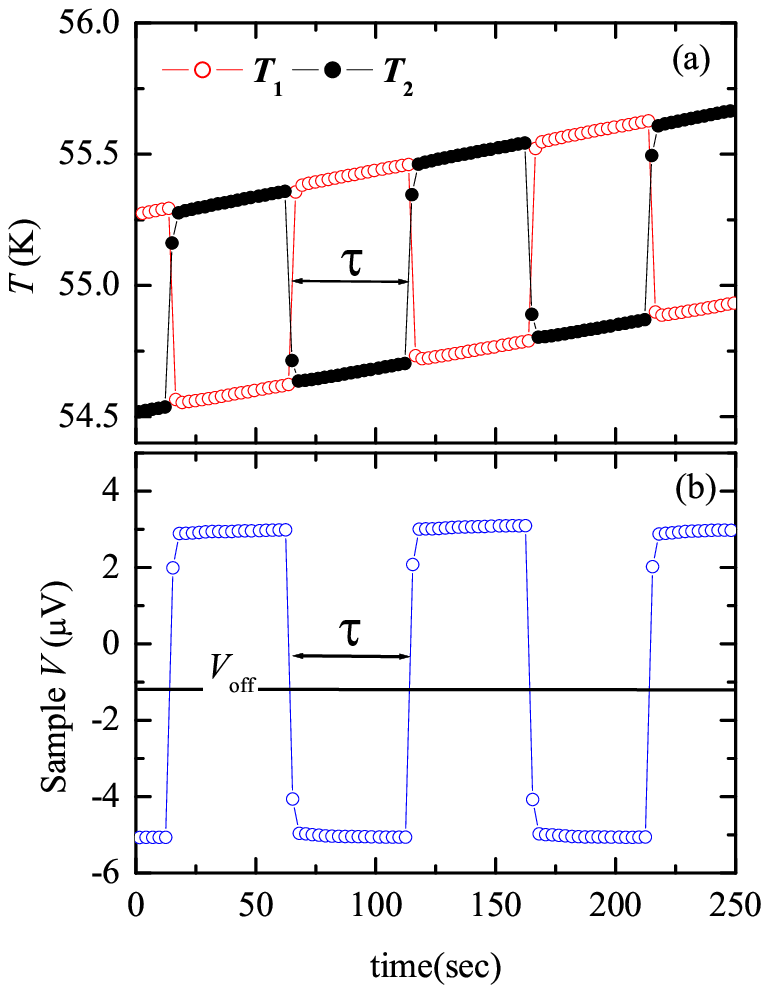}\includegraphics[width=0.5\linewidth]{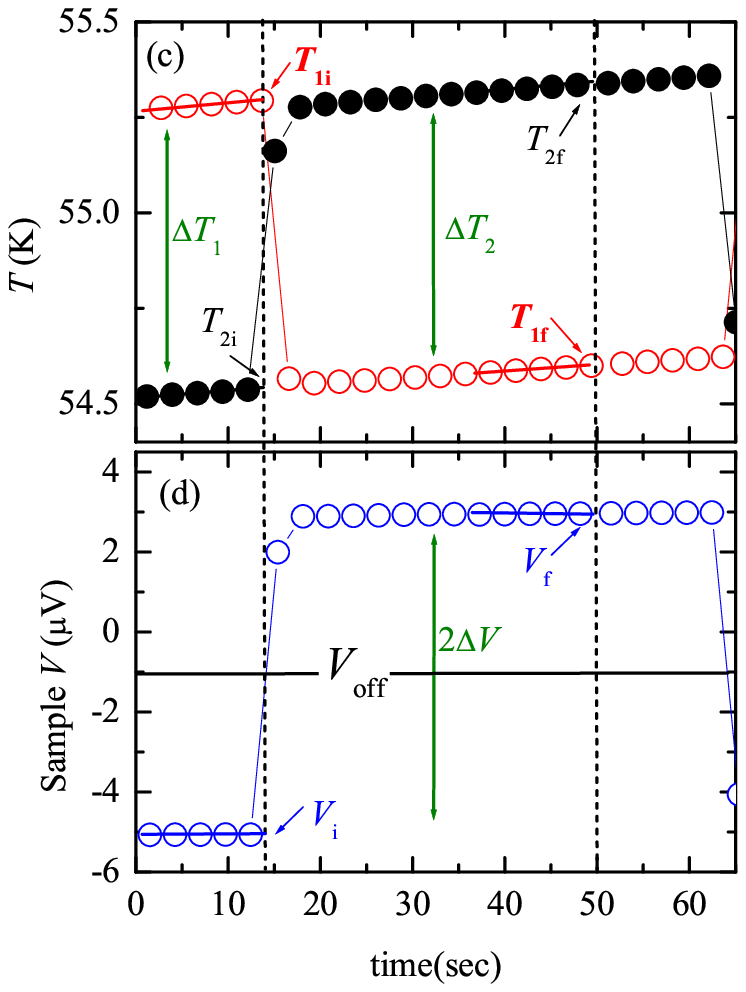}
\caption{Measurement procedure to extract the TEP from data corresponding to the measurement performed near 55 K on Pt-wire \textit{versus}
phosphor-bronze wire. Actual time period ($\tau$) between subsequent cycles, used to calculate TEP, was 50 sec. (a) Measured temperatures of both
thermometers ($T_{1}$ and $T_{2}$) and (b) sample voltage ($V$) as a function of time. Note small ($\sim$0.1 K/min) drift superimposed on data.
(c) (d) One cycle of measurement to determine parameters $\Delta T$, $\Delta V$: initial temperature $T_{i}$, final temperature $T_{f}$, initial
voltage $V_{i}$, final voltage $V_{f}$ and offset voltage $V_{off}$. The solid lines represent the linear fit to the measurement data. The
temperature difference for $T_{1}$ ($T_{2}$) is determined by $\Delta T_{1}$=$T_{1i}$-$T_{2i}$ ($\Delta T_{2}$=$T_{2f}$-$T_{1f}$) so that
2$\Delta T$=$\Delta T_{1}$+$\Delta T_{2}$. The voltage difference is calculated 2$\Delta V$=$V_{f}$-$V_{i}$ (see text).}\label{measure}
\end{figure*}

When we apply a small temperature difference across the sample, the temperatures ($T_{1}(t)$, $T_{2}(t)$) and a voltage ($V(t)$) are recorded as a
function of time, as illustrated in Fig. \ref{measure}. $T_{1}$ and $T_{2}$ are the temperatures of the two Cernox thermometers that the sample
spans. $t_{i}$ represents the time just before alternating power to the heaters (e.g. \#1 on and \#2 off) and $t_{f}$ indicates the time just
before the next power switch (e.g. \#1 off and \#2 on). As shown in Figs. \ref{measure} (c) and \ref{measure} (d) in particular, from a linear
fit of the measured voltage and temperature as a function of time, $\Delta T$ and $\Delta V$, respectively, the sample temperature $T_{av}$ and
the TEP ($S = -\Delta V/\Delta T$) are calculated using the following equations.
\begin{eqnarray}{\label{E1}}
2\Delta T &=& (T_{2f}-T_{1f})+(T_{1i}-T_{2i})\nonumber \\
2\Delta V &=& V_{f}-V_{i}\nonumber \\
T_{av}&=&\frac{(T_{2f}+T_{1f})+(T_{2i}+T_{1i})}{4}\nonumber
\end{eqnarray}
Since the temperature difference is generated by alternately applying power to one of the heaters, the measured voltage corresponds to 2$\Delta
V$. Thus, the TEP of sample is calculated by $S=-2\Delta V/2\Delta T$. Figure \ref{measure} shows the data corresponding to a measurement
performed near 55 K on a platinum (Pt) wire sample, using phosphor-bronze lead wires. The puck temperature was ramped at the rate of 0.1 K/min. A
complete cycle, used to determine $\Delta T$ and $\Delta V$, took a time period ($\tau$) of 50 sec. The parameters ($T_{1i}$, $T_{1f}$, $T_{2i}$,
$T_{2f}$, $V_{i}$ and $V_{f}$) were determined by a linear fit of the data as a function of time as shown in Fig. \ref{measure} (c) and (d).

The heater current ($I$) and time period ($\tau$), needed to generate given $\Delta T$, are not easy to estimate a prior, because of the
temperature dependence of multiple parameters, such as the thermal conductivity and heat capacity of the sample, sample stage and all electrical
wiring of the apparatus. Therefore, the current and measurement time for given $\Delta T$ were determined empirically at several temperatures by
applying constant power to one of the heaters. For determining the final temperature and voltage, after switching heater from one to the other,
the number of data point for linear fit was selected within constant temperature and voltage region as a function of time. Although it depends on
the sample under investigation, typical values of $\tau\sim$ 45 sec at 2 K and $\tau\sim$ 150 sec at 300 K for this setup allowed an accurate
determination of the final values of $T_{f}$ and $V_{f}$. Typical values of the heater current were $I\sim$ 0.8 mA to generate $\Delta T\sim$ 0.2
K at 2 K, and $I\sim$ 5 mA to generate $\Delta T\sim$ 1.0 K at 300 K.

By utilizing two heaters and an alternating gradient $\Delta T$, we avoid problems associated with offset voltages. $V_{i}$ and $V_{f}$ represent
the thermal voltages in the circuit, which include spurious voltages and the TEP of lead wires. In fact, for very low values of the TEP, it is
often necessary to consider an offset voltage ($V_{off}$) in the system and circuit. A common source of spurious voltage, for example, is the
wiring of the system from the voltmeter to the sample space since there is a thermal gradient and several soldering points between various wires.
We found that the value of $V_{off}$ for this setup depended on temperature; it was $\sim$0.5 $\mu$V around 300 K and $\sim$ -1.5 $\mu$V around
10 K. If we suppose that $V_{off}$ is independent of the small $\Delta T$ across the sample and has a small temperature dependence as a function
of time (adiabatic approximation) $V_{off}$ can be easily canceled out using two heaters as shown in Fig. \ref{measure} (d).

\begin{figure}
\centering
\includegraphics[width=0.5\linewidth]{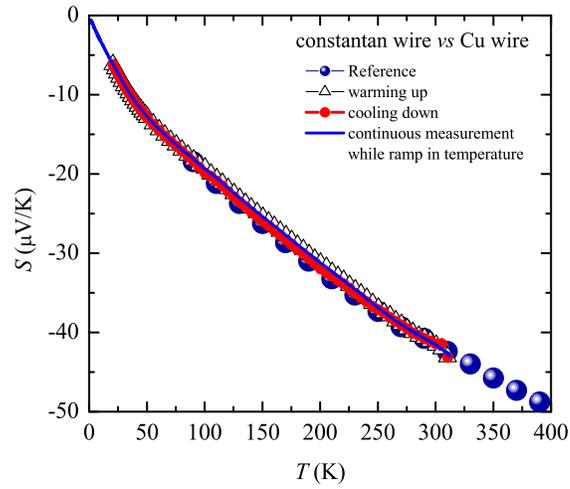}
\caption{TEP of constantan wire \textit{versus} copper wire. Warming up and cooling down indicate the measurement data using the stable
temperature method. The solid line shows the TEP values using the alternating heating method by slowly drifting system temperature. The detailed
explanations are in the text. We used the reference data provided from MMR Technologies with constantan as a standard.}\label{Constantan}
\end{figure}

In the early stage of testing this measurement setup, the process of collecting data was checked by measuring the constantan wire (100 $\mu$m
diameter) against copper wire ($\sim$ 20 $\mu$m diameter). Since constantan wire has been known to have large TEP value compared to copper wire,
the system can be tested without correcting the contribution of copper wire as shown in Fig. \ref{Constantan}. In this test running, we used the
following two protocols. Firstly, a stable temperature method was applied; in this measurement the sample puck was held at a constant temperature
and the TEP of the constantan wire using either one heater or two heaters was measured and found to be basically same within error bar of this
measurement setup. However, the TEP data for the constantan wire showed a small hysteresis upon cooling and warming between 50 and 260 K with a
maximum difference of about 2 \%. The origin of this hysteresis is not clear, we expect this that it is based on different relaxation times to
stablize the temperatures of the system.

Secondly we adopted an alternate method which was to measure the TEP while slowly warming the system temperature with the ramp rate of 0.1 K/min
below 10 K and of 0.45 K/min above 100 K (shown for a measurement of Pt wire in Fig. \ref{measure} (a) for $T\sim$55 K). As temperature increases
higher than 10 K, the ramp rate was increased for certain temperature range, for instance 0.2 K/min up to 20 K and 0.3 K/min up to 100 K. It is
worth noting that if the system temperature is slowly warming, it is necessary to carefully consider the time dependence of the sample
temperatures and voltages. In this case we calculated $\Delta T$ and $\Delta V$ from a linear fit of the data. Continuous measurements while
ramping temperature provide a high density of data and reduce the measurement time. In general it takes 16 hours to run from 2 to 350 K. This is
in contrast to our finding that the relaxation time to stablize a sample stage completely under high vacuum at a single temperature is longer
than one hour. Figure \ref{Constantan} shows the TEP of Constantan wire based on these two protocols. In this test run the agreement between
measured results and the reference data \cite{Constantan} is reasonable. The TEP extracted by the second protocol (slow drift of the system
temperature) lies between the data taken on warming and cooling using the stable temperature method.

\section{System Calibration and Sample TEP}

Since the wires attached to the sample are either copper or phosphor-bronze, a second thermal voltage is also generated. The measured TEP is then
\begin{eqnarray}{\label{E1}}
S_{measured}=S_{sample} - S_{wire}
\end{eqnarray}
Here $S_{wire}$ represents the sum of the wire and all system contributions. When measuring an unknown sample the TEP is then the sum of
$S_{wire}$ and $S_{measured}$.

\begin{figure}
\centering
\includegraphics[width=0.5\linewidth]{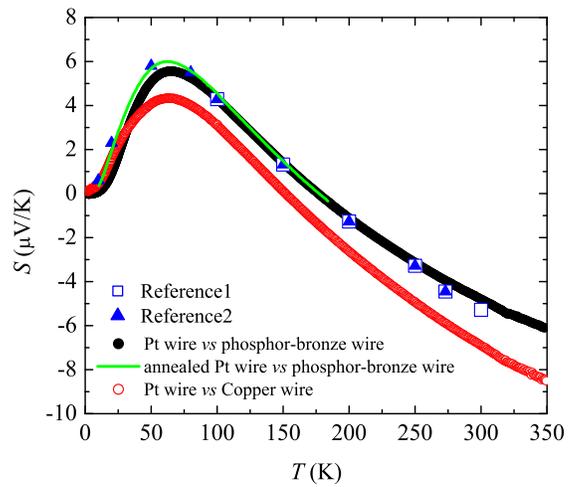}
\caption{TEP of Pt-wire \textit{versus} phosphor-bronze wire and Pt-wire \textit{versus} copper wire. Circles and solid line represent the
measured data from this work without any corrections. Both reference 1 (open squares) and reference 2 (solid triangles) data are from Ref.
\cite{Blatt}.}\label{Ptwire}
\end{figure}

\begin{figure*}
\centering
\includegraphics[width=0.5\linewidth]{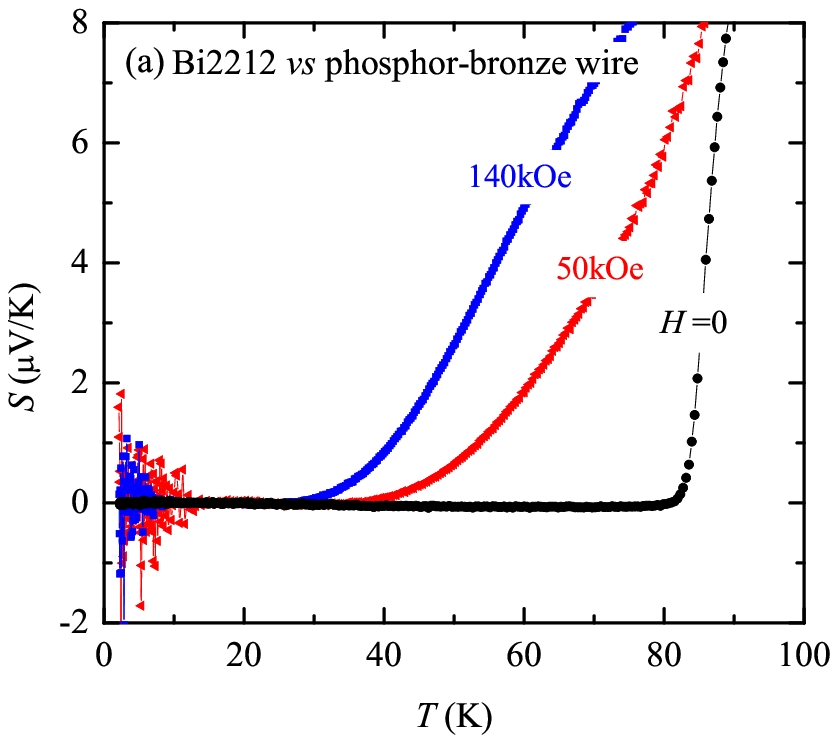}\includegraphics[width=0.5\linewidth]{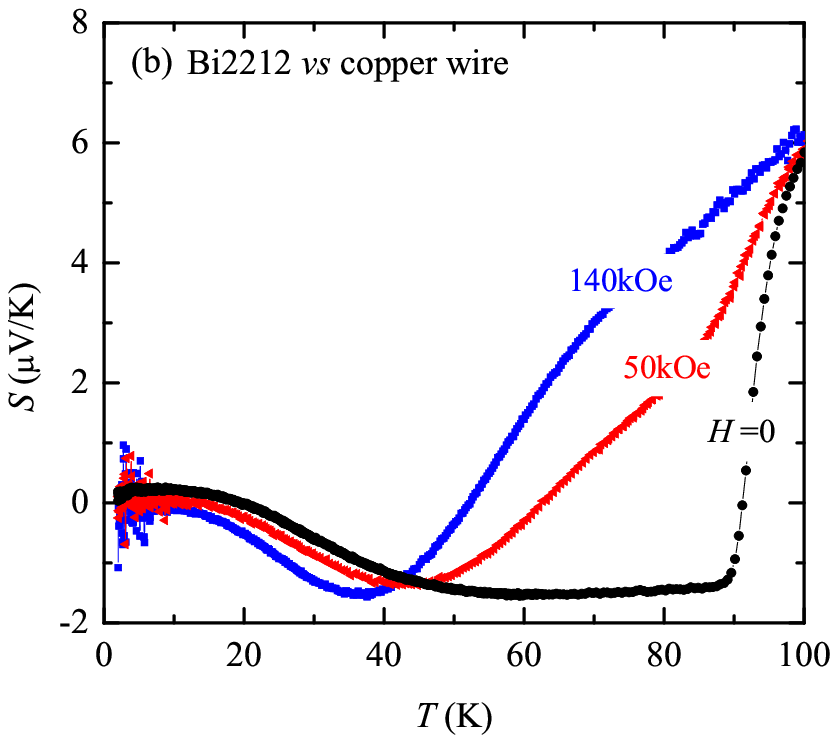}
\caption{Calibration measurements of lead wires. (a) TEP of Bi2212 \textit{versus} phosphor-bronze wire and (b) Bi2212 \textit{versus} copper
wire as a function of temperature at several constant magnetic fields.}\label{Bi2212}
\end{figure*}

The TEP of copper is strongly dependent on magnetic impurities below 100 K due to the Kondo effect \cite{Blatt} and therefore no reliable (or
universal) reference data set is available for low temperatures. On the other hand, a superconducting material is a suitable reference because
$S$ = 0 in superconducting state. In the present study Pt-wire and Bi2212 high $T_{c}$ superconductors were each, separately, mounted between the
two sample stages and calibration measurements were performed. These were sufficient for determining the lead wire contribution $S_{wire}$. For
the high temperature region pure Pt-wire ($\sim$50 $\mu$m diameter) was used as a reference. Figure \ref{Ptwire} shows the TEP of the Pt-wire
\textit{versus} copper wire and Pt-wire \textit{versus} phosphor-bronze wire. The result of Pt-wire \textit{versus} phosphor-bronze wire is in
good agreement with the absolute TEP value of Pt \cite{Blatt} which implies that the absolute TEP value of phosphor-bronze wire is negligible.
Note that below 100 K the Pt-wire manifests slightly different TEP responses depending on the heat treatment (annealing) of wire. At low
temperatures we employed two superconducting Bi2212 compounds with $T_{c}$ about $\sim$82 K and $\sim$92 K, where the different $T_{c}$ values may
be due to the heating of sample in air. The results of the TEP measurement for Bi2212 against copper and phosphor-bronze wire are shown in Fig.
\ref{Bi2212}. In this calibration measurement samples were mounted on the two sample stages with GE-varnish. The copper and phosphor-bronze wire
were attached to the sample using silver epoxy (top configuration of Fig. \ref{holder} (c)). Here we used Bright Brushing Gold to attach the wire
to the Bi2212 because using only silver epoxy provided a poor electrical contact, usually on the order of $10^{3}$ $\Omega$. After painting on
the Bright Brushing Gold, the sample was heated up to 400 $^{o}$C quickly, held for 5 min and air quenched to room temperature, where the contact
resistance was reduced to below 100 $\Omega$.

\begin{figure*}
\centering
\includegraphics[width=0.5\linewidth]{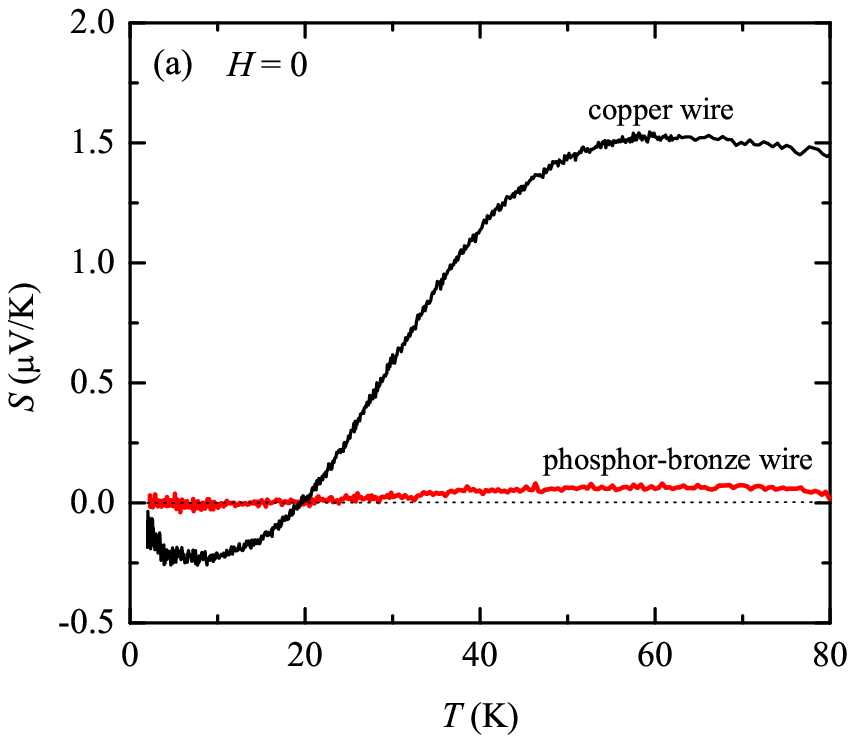}\includegraphics[width=0.5\linewidth]{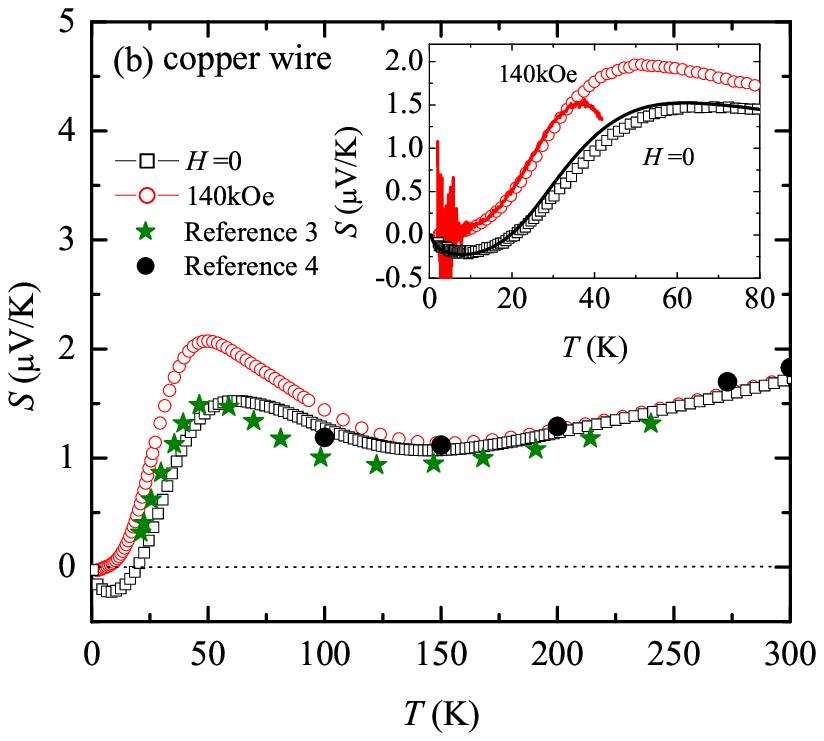}
\caption{(a) Absolute TEP of copper and phosphor-bronze wire below 80K. The data are taken from Fig. \ref{Bi2212}. (b) Calibrated TEP curve of
copper wire at $H$=0 (open square) and 140 kOe (open circle). Both closed circles (reference 3) and stars (reference 4) were taken from Ref.
\cite{Blatt}. Inset: expanded view for low temperature range. The symbols present the measured TEP of copper wire against to phosphor-bronze wire.
No correction was added. Solid lines are taken from Fig. \ref{Bi2212} (b).}\label{copper}
\end{figure*}

The absolute TEP of copper and phosphor-bronze wire we measured and of copper, from the literature, is shown in Fig. \ref{copper}. Because $S$ =
0 in the superconducting state, the observed TEP is the absolute TEP of copper and phosphor-bronze wire. From Fig. \ref{copper} (a) it is
dramatically clear that the absolute TEP value of phosphor-bronze wire is very small, $S\ll$ 0.5 $\mu$V/K, up to 80 K. For copper wire the
agreement between measured results and the literature data is reasonable. The inset of Fig. \ref{copper} (b) shows the low temperature TEP of
copper wire. For the copper wire measured against phosphor-bronze, no correction was added. These data indicate a fairly good agreement with the
data taken from Fig. \ref{Bi2212} (b). The estimated uncertainty for the copper wire is about 0.3 $\mu V$/K. In addition to the subtraction
errors, we believe that this disagreement is due to a difference in quality of the copper wire in Ref. \cite{Blatt} and that used in this
measurement.

As an aside, it should be noted that the low temperature, oscillatory behavior of the Bi2212 sample for $H>$0 (Fig. \ref{Bi2212}) is reproducible.
Although similar behavior was observed in the Nernst signal and associated with the plastic flow of the vortices \cite{Wang2}, the origin of this
phenomena is still somewhat unclear.

Previous TEP measurements at low temperatures and in high magnetic fields have had to take into account the significant contribution of background
voltage. By using well-known elemental metal wires of copper or gold and superconducting materials, these background contributions can be
accounted for, correcting the background contribution. For small single crystals an alternating AC current technique, utilizing a thermocouple,
has been used to measure TEP under high magnetic fields for a wide range of temperatures \cite{Choi, Resel}. Although the thermocouple wire
provides a good sensitivity for relative temperatures, an accurate determination of $\Delta T$ in high magnetic fields becomes difficult and large
efforts are needed to calibrate the field dependence of the thermocouple wire.

In order to exclude the difficulties due to the magneto-thermoelectric power (MTEP) measurement based primarily on the field dependence of
$S_{wire}$ and thermometer calibrations, we selected phosphor-bronze wire and Cernox. Whereas the TEP of copper (Cu) wire is not small and shows a
field dependence, phosphor-bronze wire provides essentially zero TEP over wide temperature range and is almost temperature and field independent
\cite{Wang3} as shown in Figs. \ref{Bi2212} and \ref{copper}. Therefore, in this measurement setup the magnetic field dependence of TEP of
samples, including the quantum oscillation (de Haas-van Alphen oscillation) at low temperatures, can be reliably measured.

\begin{figure*}
\centering
\includegraphics[width=0.5\linewidth]{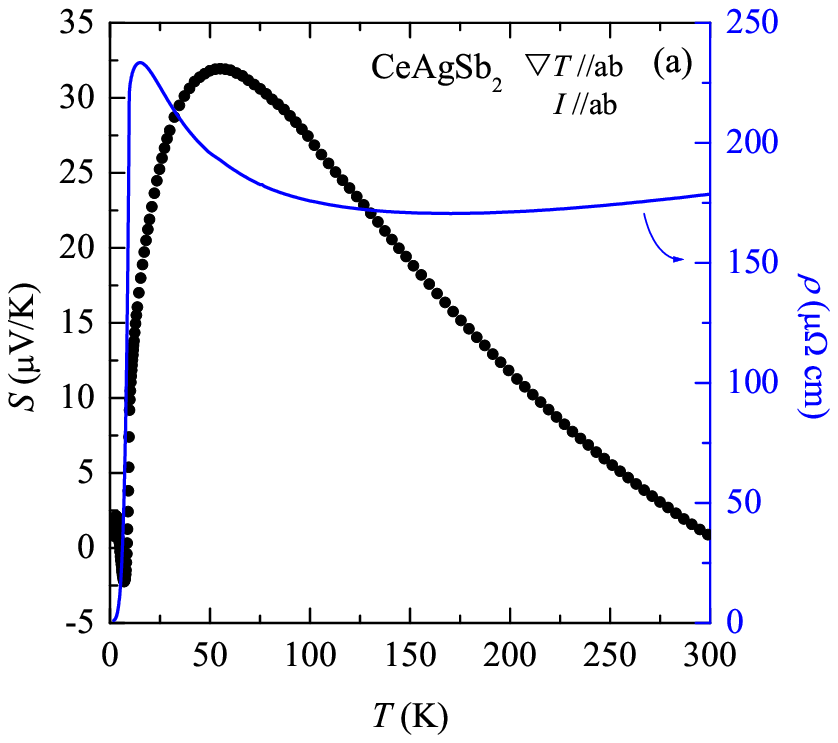}\includegraphics[width=0.5\linewidth]{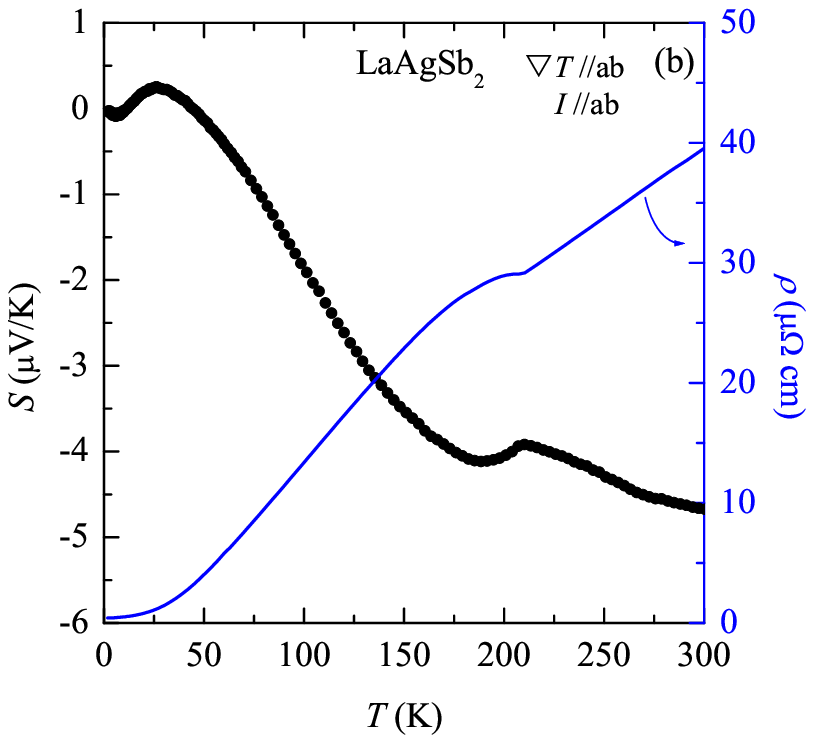}
\caption{TEP (left axis) and electrical resistivity (right axis) as a function of temperature between 2 and 300 K for (a) CeAgSb$_{2}$ and (b)
LaAgSb$_{2}$. Both the temperature gradient and electrical current are applied in the tetragonal $ab$-plane.}\label{CeLaAgSb1}
\end{figure*}

\begin{figure*}
\centering
\includegraphics[width=0.5\linewidth]{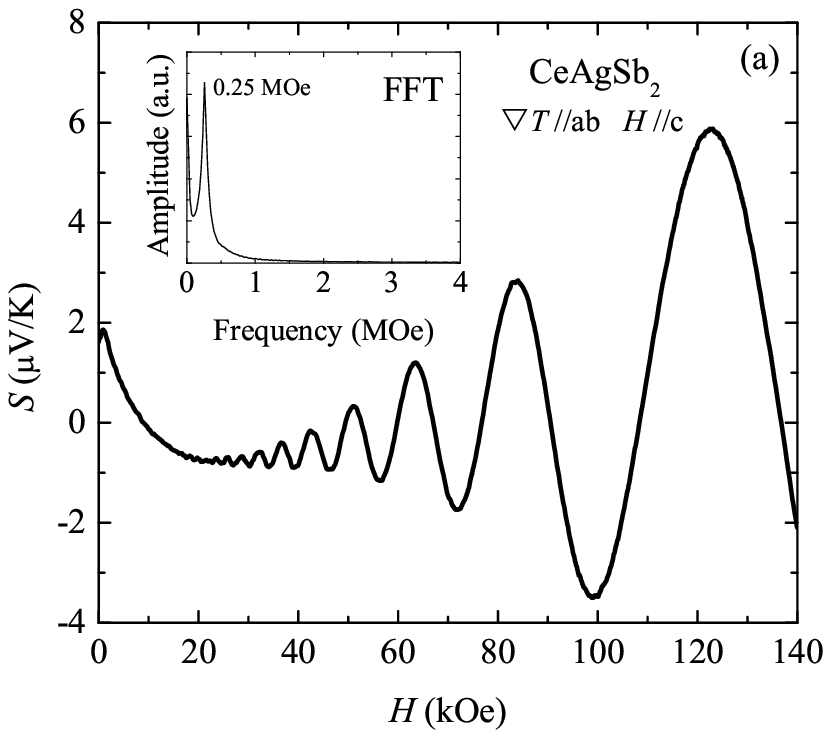}\includegraphics[width=0.5\linewidth]{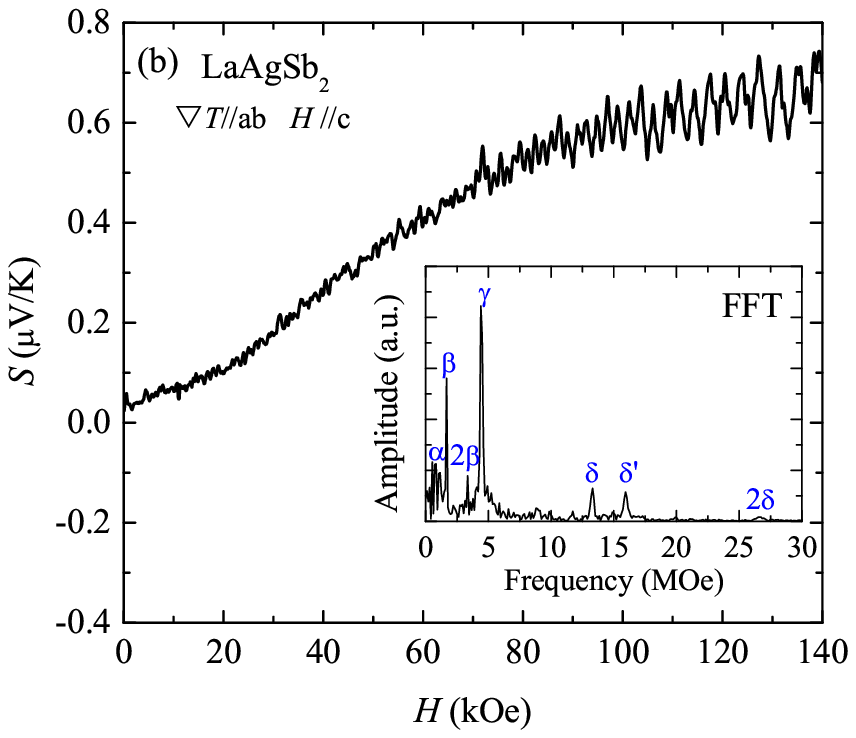}
\caption{Magneto-thermoelectric power of LaAgSb$_{2}$ and CeAgSb$_{2}$ at 2.3 K for $H\parallel c$. Labels for frequencies shown in the inset of
(b) are taken from Ref. \cite{Myers2}.} \label{CeLaAgSb2}
\end{figure*}

To demonstrate the versatility and reliability of this technique two research samples (as opposed to wires of Cu or Pt) were measured, the TEP
data are shown in Fig. \ref{CeLaAgSb1} as a function of temperature and Fig. \ref{CeLaAgSb2} as a function of applied magnetic field. LaAgSb$_{2}$
has been observed to have a charge density wave order at $\sim$210 K and $\sim$185 K \cite{Myers1, Song}, and CeAgSb$_{2}$ was characterized as a
ferromagnetic Kondo lattice compound with Curie temperature $T_{c}$=9.8 K \cite{Myers1}. In both compounds de Haas-van Alphen (dHvA) oscillations
at low temperatures have been observed \cite{Myers2}. These single crystals were grown by excess Sb flux \cite{Myers1}. Samples were prepared with
dimensions about 0.8$\times$0.2$\times$2.5 mm$^{3}$ for LaAgSb$_{2}$ and 0.8$\times$0.2$\times$3 mm$^{3}$ for CeAgSb$_{2}$. Zero field
measurement of resistivity and TEP of both materials are presented in Fig. \ref{CeLaAgSb1}. The resistivity data are consistent with earlier
study and the TEP has clear features at the same transition temperatures. For $H\parallel c$ at 2.3 K with $\triangle T$=0.2 K, for both
materials, dHvA type oscillations were observed in TEP as a function of field, $S(H)$, shown in Fig. \ref{CeLaAgSb2}. Fourier analysis (fast
Fourier transform) of the $S(H)$ data reveals peaks in the spectrum. The observed frequencies match the frequencies obtained from resistivity and
magnetization \cite{Myers2}. The detailed data analysis will be published elsewhere \cite{mun}. So as to provide a clear sense of how readily
$S(T, H)$ data can be aquired using this technique, it should be noted that the temperature dependence of TEP was taken over $\sim$14 hours and
the field dependence was taken with the ramp rate of 25 Oe/sec ($\sim$2 hours).

The accuracy of this technique was estimated by using the measurement of Pt and Cu wire. The estimated uncertainty of this system over all
temperature ranges falls within a maximum $\pm$1 $\mu$V/K, and the relative accuracy is within a maximum of 10 \%. In the high temperature
region, roughly above 100 K, the main uncertainty originates from inaccurate determination of the $\Delta T$ due to the relatively low
sensitivity of the Cernox. The absolute and relative temperature of Cernox was observed within a resolution of 4 mK at low temperatures, the
relative error at high temperatures falls within $\sim$ 200 mK. For materials having low thermal conductivity, the error may be larger due to the
temperature difference between sample and thermometer. For materials having small TEP, less than 0.5 $\mu V$/K, the error can also be larger due
to noise. More contributions to the error need to be considered for TEP measurements in the magnetic field. For instance, due to the heat
conducting environment which is mainly caused by induced current by applying magnetic fields ($d\Phi/dt$), it is very important to make sure that
the ramp rate of magnetic field should be slow enough to avoid additional heating and reduce the induced voltage due to the open loop.
Alternatively, the TEP can be measured stepping the magnetic field with the magnet in persistent mode for each value of the field.

\section{Summary of Technical Parameters and Reference Information}

\begin{itemize}
\item  Operation range: temperature range from 2 to 350 K and magnetic fields up to 140 kOe.

\item  Limit of sample dimension: the length of sample is longer than 1.5 mm (smaller than this length has not been tested).

\item  $\Delta T$: from 0.1 to 2.5 K, depending on the temperature and the absolute TEP value of sample.

\item Ramp rate of system temperature: it can be varied up to 1 K/min.
For example, in the calibration measurement, it was selected 0.1 K/min up to 10 K, 0.35 K/min up to 100 K and 0.45 K/min above 100 K.

\item Estimated accuracy: maximum of $\pm$±1 $\mu$V/K and 10\% depending on the temperature and sample. The limit of accuracy is mainly due
to the uncertainty of the thermometer and the thermal contact between the sample and the thermal stage. If the absolute TEP of the sample is
smaller than 0.5 $\mu$V/K the fluctuation of the sample voltage was observed.
\\

\item Copper wire: 0.025 mm diameter, Puratronic, 99.995\% (metals basis), Alfa Aesar. Detected impurity elements are Fe, Ag, O, S (as provided by supplier).

\item Phosphor-Bronze wire: Cu$_{0.94}$Sn$_{0.06}$ alloy, 0.025 mm dia, GoodFellow.

\item Platinum wire: 0.05 mm diameter, 99.95\% (metals basis), Alfa Aesar.

\item Silver epoxy: H20E, Epotek.

\item Strain gauge : FLG-02-23, 0.2$\times$1.4 mm$^2$ grid made by Cu-Ni alloy and 3.5$\times$2.5 mm$^2$ thin epoxy backing, Tokyo Sokki Kenkyujo Co.,
Ltd.

\item Silver paste: DuPont 4929N silver paint, DuPont, Inc.

\item Stycast 1266: Emerson \& Cuming, Inc.

\end{itemize}

\ack We would like to thank A. Kaminski for providing Bi2212 samples, J. Frederick and S. A. Law for preparing samples RAgSb$_{2}$ and M. E.
Tillman, A. Kreyssig, M. D. Vannette, C. Martin and M. A. Tanatar for valuable discussion to this project. C$_{8}$H$_{10}$N$_{4}$O$_{2}$ for this
work was provided, in part, by C. Petrovic. Work at Ames Laboratory was supported by the Basic Energy Sciences, U.S. Department of Energy under
Contract No. DE-AC02-07CH11358. Milton S. Torikachvili gratefully acknowledges support of the National Science Foundation under DMR-0805335.

\clearpage

\clearpage

\clearpage

\clearpage

\clearpage

\clearpage

\clearpage

\clearpage

\end{document}